\documentclass{bmcart}

\usepackage{amsthm,amsmath}
\usepackage[utf8]{inputenc} %unicode support

\usepackage{enumitem}
\usepackage{amssymb}
\usepackage{mathrsfs}
\usepackage{graphicx}
\usepackage{cite}
\usepackage{bbm}
\usepackage{caption}

\startlocaldefs
\endlocaldefs

\begin{document}

%%% Start of article front matter
\begin{frontmatter}

\begin{fmbox}
%\dochead{Vision Paper}


%\title{6G Wireless: Smart Radio Environments Enabled by Reconfigurable Intelligent (Meta-) Surfaces and Wireless AI}
%\title{On Wireless Futures: Smart Radio Environments Empowered by Reconfigurable Intelligent (Meta-) Surfaces}
\title{Smart Radio Environments Empowered by}
\title{AI Reconfigurable Meta-Surfaces:}
\title{An Idea Whose Time Has Come}

%%%%%%%%%%%%%%%%%%%%%%%%%%%%%%%%%%%%%%%%%%%%%%
%%                                          %%
%% Enter the authors here                   %%
%%                                          %%
%% Specify information, if available,       %%
%% in the form:                             %%
%%   <key>={<id1>,<id2>}                    %%
%%   <key>=                                 %%
%% Comment or delete the keys which are     %%
%% not used. Repeat \author command as much %%
%% as required.                             %%
%%                                          %%
%%%%%%%%%%%%%%%%%%%%%%%%%%%%%%%%%%%%%%%%%%%%%%

\author[
   addressref={aff1},                   % id's of addresses, e.g. {aff1,aff2}
   corref={aff1},                       % id of corresponding address, if any
   %noteref={n1},                        % id's of article notes, if any
   email={marco.direnzo@l2s.centralesupelec.fr}   % email address
]{\inits{MDR}\fnm{Marco} \snm{Di Renzo}}
\author[
   addressref={aff2},
   email={merouane.debbah@huawei.com}
]{\inits{TBA}\fnm{Merouane} \snm{Debbah}}
\author[
   addressref={aff3},
   email={dinhthuy.phanhuy@orange.com}
]{\inits{TBA}\fnm{Dinh-Thuy} \snm{Phan-Huy}}
\author[
   addressref={aff4},
   email={alessio.zappone@l2s.centralesupelec.fr}
]{\inits{TBA}\fnm{Alessio} \snm{Zappone}}
\author[
   addressref={aff5},
   email={slim.alouini@kaust.edu.sa}
]{\inits{TBA}\fnm{Mohamed-Slim} \snm{Alouini}}
\author[
   addressref={aff6},
   email={yuenchau@sutd.edu.sg}
]{\inits{TBA}\fnm{Chau} \snm{Yuen}}
\author[
   addressref={aff7},
   email={vincenzo.sciancalepore@neclab.eu}
]{\inits{TBA}\fnm{Vincenzo} \snm{Sciancalepore}}
\author[
   addressref={aff8},
   email={alexandg@di.uoa.gr}
]{\inits{TBA}\fnm{George C.} \snm{Alexandropoulos}}
\author[
   addressref={aff9},
   email={jakob.hoydis@nokia-bell-labs.com}
]{\inits{TBA}\fnm{Jakob} \snm{Hoydis}}
\author[
   addressref={aff10},
   email={haris.gacanin@nokia-bell-labs.com}
]{\inits{TBA}\fnm{Haris} \snm{Gacanin}}
\author[
   addressref={aff11},
   email={julien.derosny@espci.fr}
]{\inits{TBA}\fnm{Julien} \snm{de Rosny}}
\author[
   addressref={aff12},
   email={ahcene.bounceur@univ-brest.fr}
]{\inits{TBA}\fnm{Ahcene} \snm{Bounceu}}
\author[
   addressref={aff13},
   email={g.lerosey@gmail.com}
]{\inits{TBA}\fnm{Geoffroy} \snm{Lerosey}}
\author[
   addressref={aff11},
   email={mathias.fink@espci.fr}
]{\inits{TBA}\fnm{Mathias} \snm{Fink}}

%%%%%%%%%%%%%%%%%%%%%%%%%%%%%%%%%%%%%%%%%%%%%%
%%                                          %%
%% Enter the authors' addresses here        %%
%%                                          %%
%% Repeat \address commands as much as      %%
%% required.                                %%
%%                                          %%
%%%%%%%%%%%%%%%%%%%%%%%%%%%%%%%%%%%%%%%%%%%%%%

\address[id=aff1]{%
  \orgname{Laboratoire des Signaux et Syst\`emes, CNRS, CentraleSupelec, Univ Paris-Sud, Universit\'e Paris-Saclay},
  \street{Plateau de Saclay},
  \postcode{91192}
  \city{Gif-sur-Yvette},
  \cny{France}                          % country
}
\address[id=aff2]{%
  \orgname{Mathematical and Algorithmic Sciences Lab, Huawei France R\&D},
  \street{20 Quai du Point du Jour},
  \postcode{92100}
  \city{Boulogne-Billancourt},
  \cny{France}
}
\address[id=aff3]{%
  \orgname{Orange Labs},
  \street{44 avenue de la Republique},
  \postcode{92326}
  \city{Chatillon},
  \cny{France}
}
\address[id=aff4]{%
  \orgname{Laboratoire des Signaux et Syst\`emes, CentraleSupelec, LANES Group},
  \street{Plateau de Saclay},
  \postcode{91192}
  \city{Gif-sur-Yvette},
  \cny{France}
}
\address[id=aff5]{%
  \orgname{King Abdullah University of Science and Technology (KAUST)},
  \street{Mail Box 1282},
  \postcode{23955-6900}
  \city{Thuwal},
  \cny{Kingdom of Saudi Arabia}
}
\address[id=aff6]{%
  \orgname{Singapore University of Technology and Design (SUTD)},
  \street{8 Somapah Rd},
  \postcode{487372}
  \city{Singapore},
  \cny{Singapore}
}
\address[id=aff7]{%
  \orgname{NEC Laboratories Europe},
  \street{Kurfursten-Anlage, 36},
  \postcode{69115}
  \city{Heidelberg},
  \cny{Germany}
}
\address[id=aff8]{%
  \orgname{National and Kapodistrian University of Athens},
  \street{Panepistimiopolis Ilissia},
  \postcode{15784}
  \city{Athens},
  \cny{Greece}
}
\address[id=aff9]{%
  \orgname{Nokia Bell Labs},
  \street{7 Route de Villejust},
  \postcode{91620}
  \city{Nozay},
  \cny{France}
}
\address[id=aff10]{%
  \orgname{Nokia Bell Labs},
  \street{Copernicuslaan 50},
  \postcode{2018}
  \city{Antwerp},
  \cny{Belgium}
}
\address[id=aff11]{%
  \orgname{Institut Langevin, ESPCI Paris},
  \street{1 rue Jussieu},
  \postcode{75238}
  \city{Paris},
  \cny{France}
}
\address[id=aff12]{%
  \orgname{University of Brest},
  \street{20, Avenue Victor Le Gorgeu},
  \postcode{29238}
  \city{29238},
  \cny{France}
}
\address[id=aff13]{%
  \orgname{Greenerwave},
  \street{6 Rue Jean Calvin},
  \postcode{75005}
  \city{Paris},
  \cny{France}
}

%%%%%%%%%%%%%%%%%%%%%%%%%%%%%%%%%%%%%%%%%%%%%%
%%                                          %%
%% Enter short notes here                   %%
%%                                          %%
%% Short notes will be after addresses      %%
%% on first page.                           %%
%%                                          %%
%%%%%%%%%%%%%%%%%%%%%%%%%%%%%%%%%%%%%%%%%%%%%%

%\note{Sample of title note}     % note to the article
%\note[id=n1]{Equal contributor} % note, connected to author

\end{fmbox}% comment this for two column layout

%%%%%%%%%%%%%%%%%%%%%%%%%%%%%%%%%%%%%%%%%%%%%%
%%                                          %%
%% The Abstract begins here                 %%
%%                                          %%
%% Please refer to the Instructions for     %%
%% authors on http://www.biomedcentral.com  %%
%% and include the section headings         %%
%% accordingly for your article type.       %%
%%                                          %%
%%%%%%%%%%%%%%%%%%%%%%%%%%%%%%%%%%%%%%%%%%%%%%

\begin{abstractbox}

\begin{abstract}
Future wireless networks are expected to constitute a distributed intelligent wireless communications, sensing, and computing platform, which will have the challenging requirement of interconnecting the physical and digital worlds in a seamless and sustainable manner.
Currently, two main factors prevent wireless network operators from building such networks: 1) the lack of control of the wireless environment, whose impact on the radio waves cannot be customized, and 2) the current operation of wireless radios, which consume a lot of power because new signals are generated whenever data has to be transmitted.
In this paper, we challenge the usual ``more data needs more power and emission of radio waves'' status quo, and motivate that future wireless networks necessitate a smart radio environment: A transformative wireless concept, where the environmental objects are coated with artificial thin films of electromagnetic and reconfigurable material (that are referred to as intelligent reconfigurable meta-surfaces), which are capable of sensing the environment and of applying customized transformations to the radio waves. Smart radio environments have the potential to provide future wireless networks with uninterrupted wireless connectivity, and with the capability of transmitting data without generating new signals but recycling existing radio waves.
We will discuss, in particular, two major types of intelligent reconfigurable meta-surfaces applied to wireless networks. The first type of meta-surfaces will be embedded into, e.g., walls, and will be directly controlled by the wireless network operators via a software controller in order to shape the radio waves for, e.g., improving the network coverage. The second type of meta-surfaces will be embedded into objects, e.g., smart t-shirts with sensors for health monitoring, and will backscatter the radio waves generated by cellular base stations in order to report their sensed data to mobile phones. These functionalities will enable wireless network operators to offer new services without the emission of additional radio waves, but by recycling those already existing for other purposes.
This paper overviews the current research efforts on smart radio environments, the enabling technologies to realize them in practice, the need of new communication-theoretic models for their analysis and design, and the long-term and open research issues to be solved towards their massive deployment. In a nutshell, this paper is focused on discussing how the availability of intelligent reconfigurable meta-surfaces will allow wireless network operators to redesign common and well-known network communication paradigms.
\end{abstract}

%%%%%%%%%%%%%%%%%%%%%%%%%%%%%%%%%%%%%%%%%%%%%%
%%                                          %%
%% The keywords begin here                  %%
%%                                          %%
%% Put each keyword in separate \kwd{}.     %%
%%                                          %%
%%%%%%%%%%%%%%%%%%%%%%%%%%%%%%%%%%%%%%%%%%%%%%


% MSC classifications codes, if any
%\begin{keyword}[class=AMS]
%\kwd[Primary ]{}
%\kwd{}
%\kwd[; secondary ]{}
%\end{keyword}

\end{abstractbox}
%
%\end{fmbox}% uncomment this for twcolumn layout

\end{frontmatter}

%%%%%%%%%%%%%%%%%%%%%%%%%%%%%%%%%%%%%%%%%%%%%%
%%                                          %%
%% The Main Body begins here                %%
%%                                          %%
%% Please refer to the instructions for     %%
%% authors on:                              %%
%% http://www.biomedcentral.com/info/authors%%
%% and include the section headings         %%
%% accordingly for your article type.       %%
%%                                          %%
%% See the Results and Discussion section   %%
%% for details on how to create sub-sections%%
%%                                          %%
%% use \cite{...} to cite references        %%
%%  \cite{koon} and                         %%
%%  \cite{oreg,khar,zvai,xjon,schn,pond}    %%
%%  \nocite{smith,marg,hunn,advi,koha,mouse}%%
%%                                          %%
%%%%%%%%%%%%%%%%%%%%%%%%%%%%%%%%%%%%%%%%%%%%%%

%%%%%%%%%%%%%%%%%%%%%%%%% start of article main body
% <put your article body there>

%%%%%%%%%%%%%%%%
%% Background %%
%%

\section{Wireless Futures - Beyond Communications, but Without More Power and Radio Waves}
Future wireless networks are expected be more than allowing people, mobile devices, and objects to communicate with each other \cite{1}. Future wireless networks have the potential to be turned into a distributed intelligent communications, sensing, and computing platform. Besides connectivity, more specifically, the platform will be capable of sensing the environment to realize the vision of smart living in smart cities by providing context-awareness capabilities, and of locally storing and processing information. Such processing could accommodate the time critical, ultra-reliable, and energy efficient delivery of data, and the accurate localization of people and objects in environments and scenarios where the Global Positioning System (GPS) is not an option. Future wireless networks will have to fulfill the challenging requirement of interconnecting the physical and digital worlds in a seamless and sustainable manner. \\

\textit{What is currently slowing down wireless network operators from building truly pervasive wireless networks that are capable of enabling communications, and of collecting and understanding data from the physical world}?  \\ \\

There exist two fundamental limiting factors: \\
\begin{enumerate}
  \item Wireless network operators struggle to power the continuous sensing and actuation of millions (or billions) of devices, and to continuously connect them to the Internet due to the high power consumption of the wireless interface. This originates from the current operation of wireless radios, which consume a lot of power during data communication because the radios themselves are the devices that generate every wireless signal \cite{2}. \\
  \item Wireless network operators struggle to provide users, devices, and connected objects with uninterrupted connectivity and quality of service guarantee in harsh propagation environments. This originates from the lack of control that we have of the wireless environment, whose impact on the signals cannot be adaptively customized as we desire \cite{3}. \\ \\
\end{enumerate}

The usual response of wireless network operators to the tremendous increase of traffic demands consists of using more power and emitting more radio waves. This is usually achieved by transmitting signals in new frequency bands, i.e., using more spectrum, and by deploying more cellular base stations, i.e., densifying the network. Even though new generations of wireless networks are always more energy and spectral efficient than the previous ones, the power consumption and the emission of radios waves always increase from past to new generations \cite{OrangeView}. \\ \\

Therefore, it is time to identify alternative solutions to the de facto approach ``more data via more power and more emissions of radio waves''. In this context, in particular, two intriguing questions are naturally brought to the attention of the wireless community: \\
\begin{itemize}
  \item \textit{What if wireless network operators could control the wireless environment by allowing energy-constrained devices to sense and report the measured data without using new radio waves, but by just recycling those that are generated by their own network and, possibly, without the need of batteries}? \\
  \item \textit{What if wireless network operators could customize, via a remote software-operated controller equipped with predictive capabilities, the propagation of the radio waves in the environment in order to increase the data rate without increasing the power consumption}? \\ \\
\end{itemize}

In the present paper, we put forth and elaborate on the emerging concept of smart radio environments, as the fundamental distributed wireless platform, under the control of the wireless network operators, that integrates communications, sensing, and computing capabilities, as well as the enabling technology to realize the wireless future envisioned by the two questions above. \\ \\

\textit{What is a smart radio environment}? A smart radio environment is a wireless environment that is turned into a smart reconfigurable space and that plays an active role in transferring and processing information \cite{6}. Smart radio environments largely extend the notion of software networks: Currently, the operation of wireless networks is software-controlled and elastically optimized to support heterogeneous requirements (e.g., enhanced data rate, high energy efficiency, low latency, ultra-reliability, massive connectivity of objects) \cite{4}. In our definition of smart radio environments, the wireless environment itself is turned into a software-reconfigurable entity \cite{5}, whose operation is optimized to enable uninterrupted connectivity, quality of service guarantee, and where the information is transmitted without necessarily generating new signals but recycling the existing ones whenever possible \cite{2}. \\ \\

\textit{How to tailor smart radio environments into the real world}? Fortunately, different but converging solutions are recently emerging to realize the vision of smart radio environments. This includes deploying programmable frequency-selective surfaces \cite{7}, \cite{8}, and smart reflect-arrays or mirrors \cite{9}-\cite{11} in the environment, embedding arrays of low-cost antennas \cite{6}, \cite{12}, \cite{13} into the walls of buildings, and coating the environmental objects with reconfigurable meta-surfaces \cite{14}. Meta-surfaces, in particular, are thin meta-material layers that are capable of shaping the propagation of radio waves in fully customizable ways \cite{15}, and, thus, have the potential of making the transfer and processing of information more reliable \cite{16}-\cite{IL_END}. In addition, they constitute a suitable distributed platform to perform low-energy and low-complexity sensing \cite{17}, storage \cite{18}, and analog computing \cite{IL_Computing}, \cite{19}. Thanks to these unique properties, the high controllability of the radio waves, the high deployment scalability \cite{20}, and the economic advantages that they bring about \cite{21}, reconfigurable meta-surfaces are today considered to be a core technology to fulfill the challenging requirements of future wireless networks. In the present paper, to avoid ambiguity and in agreement with the Greek etymology of the word ``meta'' (i.e., beyond), we will use the term meta-surface to denote any surface that is engineered to have properties that are not found in naturally occurring surfaces. \\ \\

\textit{Are reconfigurable meta-surfaces currently available}? We will elaborate on this question in the next sections. It suffices to say that prototypes of reconfigurable meta-surfaces are currently being developed \cite{22}, and startup companies are developing the fundamental technology that covers a wide range of the electromagnetic spectrum \cite{GreenerWave}. For example, scientists of the European-funded project VISORSURF \cite{22} have recently built the prototype of a software-controlled meta-surface that makes the wireless environment fully reconfigurable. Thanks to this breakthrough, it is today realistic to envision wireless networks where every environmental object \cite{15} is coated with an artificial thin film of electromagnetic material \cite{23}, which senses the environment and whose response to the radio waves is programmed to optimize the performance. \\ \\

At a time when the core technologies to realize reconfigurable and software-controllable meta-surfaces are considered to be feasible, communication theorists and wireless researchers are, however, challenged by three fundamental questions: \\
\begin{enumerate}
   \item \textit{How to integrate the reconfigurable meta-surfaces into wireless networks}? \\
   \item \textit{What are the ultimate performance limits of wireless networks in the presence of reconfigurable meta-surfaces}? \\
   \item \textit{How to attain such performance limits in practice}? \\ \\
 \end{enumerate}

In the next sections, we will discuss these three fundamental and open research questions, and, notably, we will propose a new communication-theoretical model that accounts for the peculiarities brought about by the smart radio environments. Furthermore, we will elaborate on tools and methods towards the theoretic and algorithmic foundation of smart radio environments. \\ \\

\section{Smart Radio Environments}
In current wireless networks, the radio environment, i.e., the physical objects that alter the propagation of the electromagnetic waves, is not controllable \cite{3}, and is perceived, in addition, as an adversary to the communication process, i.e., it has usually a negative effect that needs to be counteracted by the transmitters and receivers \cite{12}. By contrast, we define a smart radio environment as a radio environment that is turned into a smart reconfigurable space that plays an active role in transferring and processing information, and that makes more reliable the exchange of data between transmitters and receivers. \\

To better elucidate the concept of smart radio environments in the context of wireless networks, we commence this section by briefly introducing what a reconfigurable meta-surface is. Detailed information about the research efforts on designing reconfigurable meta-surfaces are reported in the sequel. Then, we discuss two examples that consider typical applications in communications and in sensing. \\

\subsection{Meta-Surfaces and Reconfigurable Meta-Surfaces}
The fundamental constituting and enabling element of the smart radio environment is the reconfigurable meta-surface. \textit{What is a meta-surface}? \\

As the Greek meaning of the word ``meta'', i.e., beyond, suggests, an electromagnetic meta-surface is a surface made of electromagnetic material that is engineered in order to exhibit properties that are not found in naturally occurring materials. A meta-surface is, in practice, an electromagnetic discontinuity, which can be defined as a complex electromagnetic structure that is typically deeply sub-wavelength in thickness, is electrically large in transverse size, and is composed of sub-wavelength scattering particles with extremely small features \cite{77}. In simple terms, a meta-surface is made of a two-dimensional array of sub-wavelength metallic or dielectric scattering particles that transform the electromagnetic waves in different ways \cite{80}. \\

\begin{figure}[!t] \label{Metasurface}
	\setlength{\captionmargin}{10.0pt}	
	\centering
	\includegraphics[width=\columnwidth]{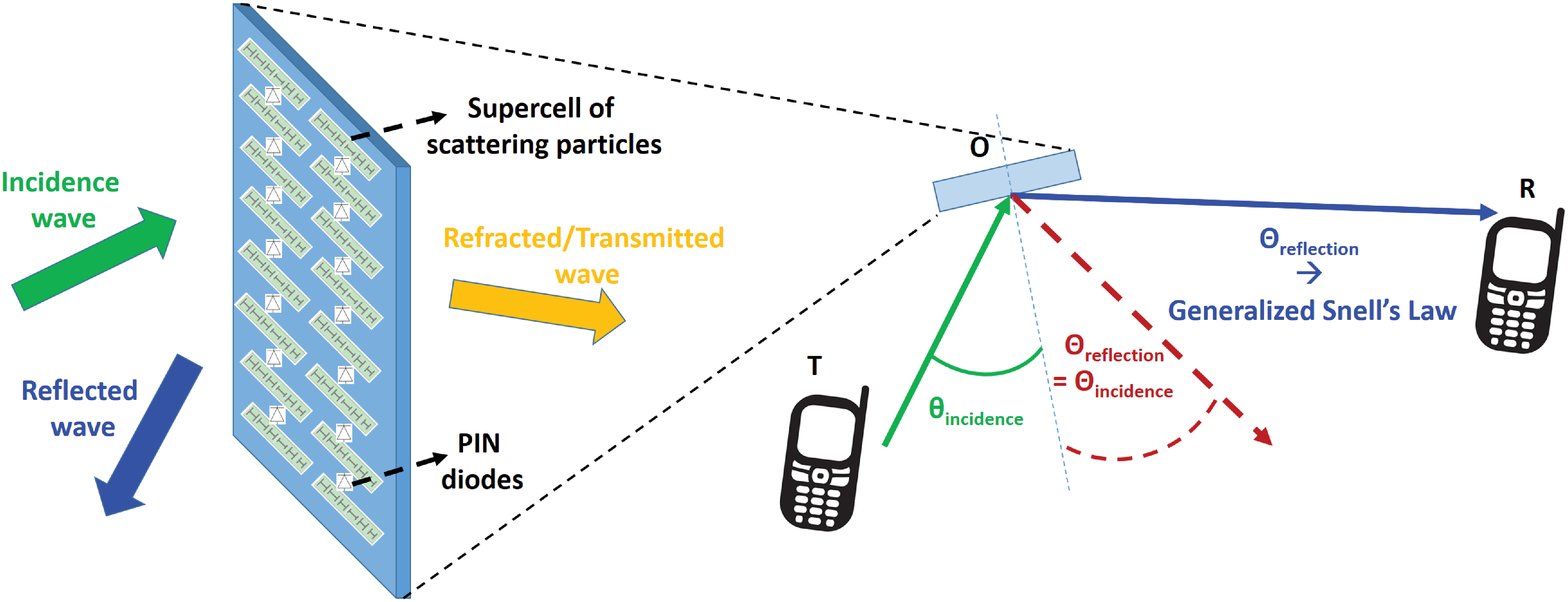}
	\caption{Working principle of reconfigurable meta-surfaces}
\end{figure}
An example of meta-surface is sketched in Fig. 1, where it is shown that it transforms an incident radio wave into a reflected radio wave and a transmitted (or refracted) radio wave. The specific arrangements of the scattering particles (e.g., full or slotted patches, straight or curved strips, various types of crosses, etc.) determine how the meta-surface transforms the incident wave into arbitrary specified reflected and transmitted radio waves \cite{165}. The major difference between a surface and a meta-surface lies in the capability of the latter of shaping the radio waves according to the generalized Snell's laws of reflection and refraction \cite{15}. For example, the angles of incidence and reflection of the radio waves are not necessarily the same in a meta-surface. A reconfigurable meta-surface is a meta-surface in which the scattering particles are not fixed and engineered at the manufacturing phase, but can be modified depending on the stimuli that the meta-surface receives from the external world. For example, multiple elementary scattering particles that realize some specific wave transformations can be connected by using electronic circuits that activate only those that synthesize the specified wave transformation of interest for a given network configuration \cite{36}. In Fig. 1, this functionality is, e.g., realized by using PIN diodes. It is worth mentioning, as better detailed in the sequel, that the reconfigurable meta-surfaces may be equipped with embedded sensors that could allow them to sense the status of the environment, e.g., the channel states between them and the base stations, and between them and the mobile terminals, and to report this information to the external world (i.e., a network controller), which is capable of configuring their operation via a feedback channel \cite{3}. \\

As discussed in \cite{80}, the synthesis and analysis of meta-surfaces (just a single meta-surface) is an extremely difficult task. An approach would consist of analyzing the meta-surfaces by using a general-purpose full-wave electromagnetic simulator. A meta-surface, however, is electrically thin (i.e., its thickness is much smaller than the wavelength), is electrically relatively large (i.e., the other dimensions are larger than the wavelength), and is composed of sub-wavelength particles with deeply sub-wavelength features. Therefore, such a brute-force approach turns out to be impractical, as it would require large memory resources and would take a prohibitive computation time, while giving little insight into the physics of the meta-surface. In the sequel, we will discuss some emerging and recent techniques to circumvent this issue, which are based on approximating the meta-surfaces as local entities of general conformal shapes \cite{84}, and in modeling the meta-surfaces as a zero-thickness sheet (also known as sheet discontinuity model) \cite{77}. \\ \\

Based on this discussion and on the complexity of modeling and optimizing just a single meta-surface, the ``fil rouge'' of the present article lies in elaborating the fundamental gaps of knowledge behind the analysis and synthesis of smart radio environments, in which many reconfigurable meta-surfaces can be deployed and need to be jointly optimized. In other words, we will address the following question: \\

\textit{If we think of smart radio environments, how to model, analyze, simulate, optimize, and orchestrate a multitude of reconfigurable meta-surfaces that are spatially distributed in a large-scale wireless network}? \\

\begin{figure}[!t] \label{NetworksToday}
	\setlength{\captionmargin}{10.0pt}	
	\centering
	\includegraphics[width=\columnwidth]{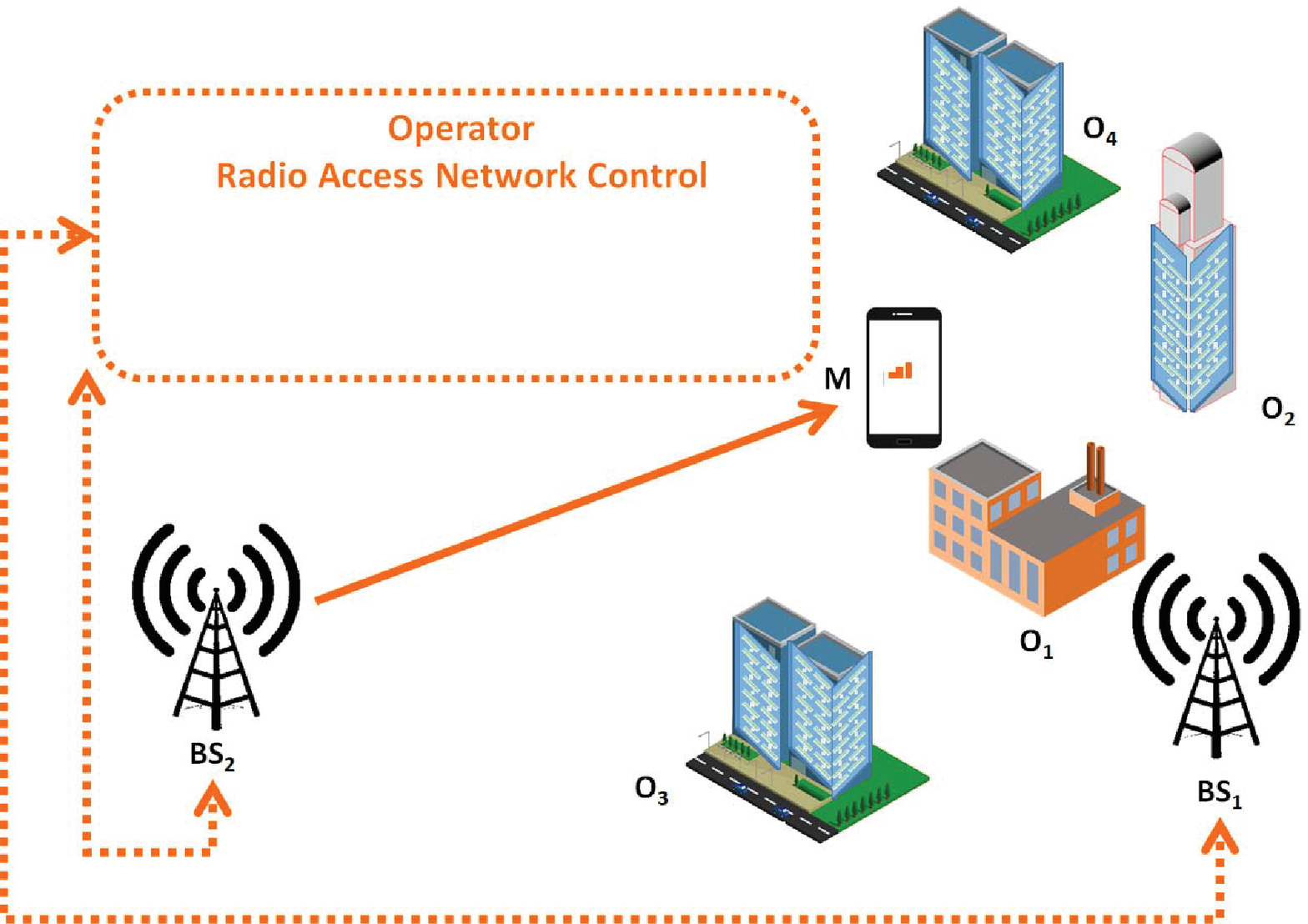}
	\caption{Current operation of wireless networks: Communications}
\end{figure}
\begin{figure}[!t] \label{NetworksTomorrow}
	\setlength{\captionmargin}{10.0pt}	
	\centering
	\includegraphics[width=\columnwidth]{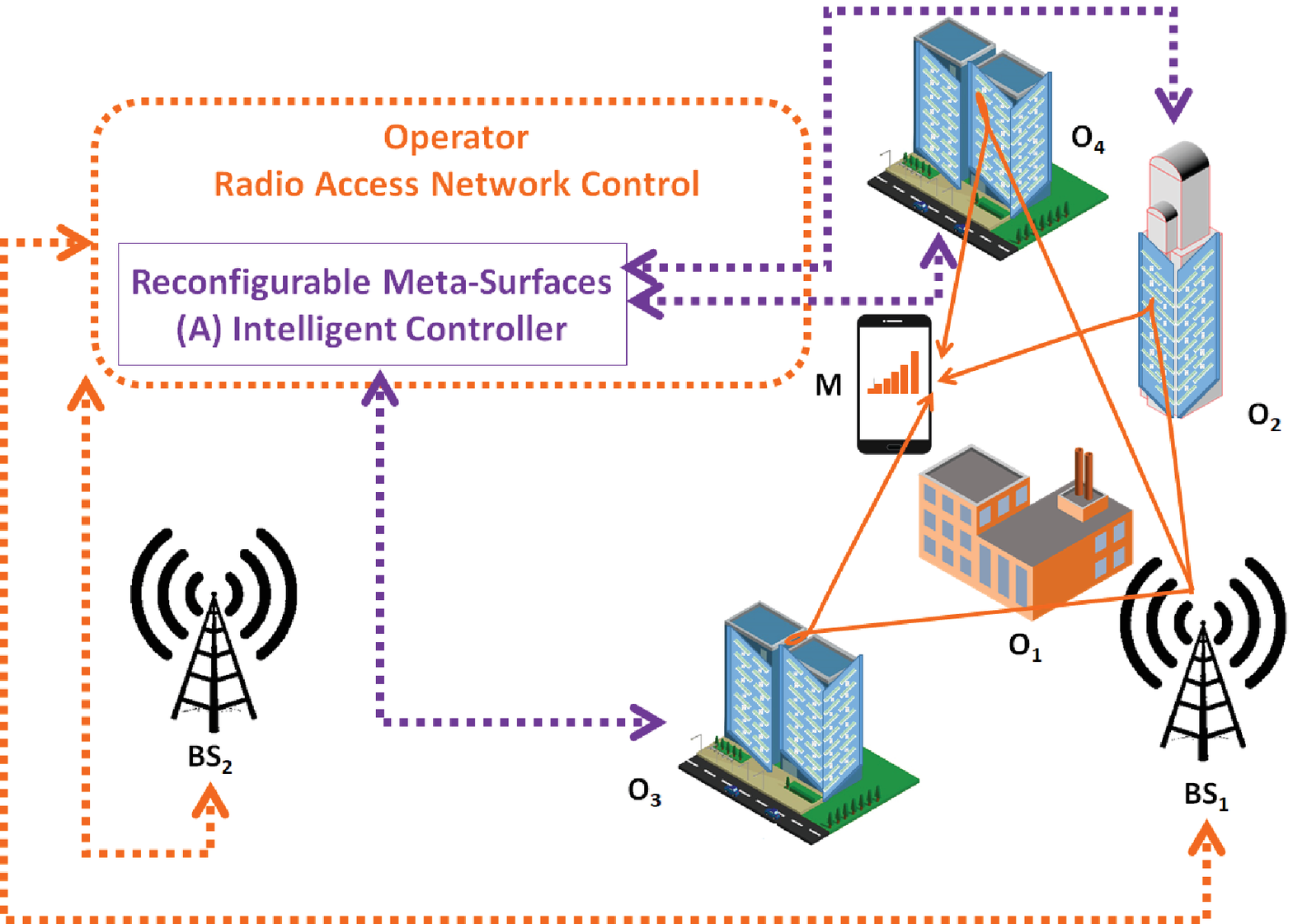}
	\caption{Smart radio environments: Communications}
\end{figure}
\subsection{Reconfigurable Environments: Improving Communications}
Figure 2 shows the operating principle that current wireless networks obey to. A mobile terminal (M) wants to connects to the Internet via a cellular network. In the absence of environmental objects (O1, O2, O3, O4), BS1 is the base station that provides the best signal to M. Due to the high blocking object O1, however, the received signal is not sufficiently strong and M connects to the Internet via BS2, while BS1 is kept active to serve other users. Since BS2 is far from M, even though it transmits at high power, the signal received by M is not sufficiently strong for high data rate transmission. \\

Let us now consider, on the other hand, Fig. 3. The objects (O1, O2, O3, O4) are now coated with intelligent (possibly AI-based, as envisioned at the end of this paper) reconfigurable meta-surfaces that modify the radio waves by introducing, in a software-controlled and programmable manner \cite{3}, localized and location-dependent gradient phase shifts onto the signals impinging upon them. Such abrupt phase discontinuity along the meta-surface is the key element for wave manipulation, e.g., to absorb, refract, reflect the signals in agreement with the generalized laws of reflection and refraction (beyond Snell's laws) \cite{15}. Figure 3 illustrates how this fundamentally changes the operation of wireless networks. The link between BS1 and M is still obstructed by the high blocking object O1. In this case, however, the responses of the intelligent reconfigurable meta-surfaces on O2, O3, and O4 are controlled and optimized to refract or reflect towards anomalous (i.e., not compliant with Snell's laws) directions, the waves throughout the network, thus altering the spatial distribution of the intended and interfering signals. For example, O2 refracts the signal from BS1, by producing a strong received signal at M, while avoiding to interfere towards other users (unwanted directions), and O3 reflects the signals towards M, thus further strengthening the intended signal at M. This is possible by capitalizing on the sensing capabilities of the intelligent reconfigurable meta-surfaces, and on their capabilities of reporting the sensed data to a network controller that processes it and computes the best wave transformations to apply in order to shape the radio waves according to the locations of the base stations and mobile terminals \cite{3}. This solution requires, in general, that the intelligent reconfigurable meta-surfaces are equipped with some power sources, e.g., batteries, energy harvesting and storage modules, or a combination of them. In addition, nano-networking protocols within the intelligent reconfigurable meta-surfaces are needed in order to support their reconfigurability \cite{72}. If the intelligent reconfigurable meta-surfaces are not equipped with sensing capabilities, the radio links can be estimated by the base stations and the mobile terminals via appropriate control signals. The base stations and the mobile terminals are then in charge of reporting this data to the network controller, which is, in turn, responsible for computing the best configuration setup of the intelligent reconfigurable meta-surfaces, and for sending the corresponding control signals to them. \\

It is worth mentioning, in addition, that the focused transmissions from the reconfigurable intelligent meta-surfaces, which are similar to time-reversal focusing \cite{RefFocusing}, could be used for enhancing the security of communication networks \cite{RefSecurity}. \\ \\

\begin{figure}[!t] \label{Backsatter}
	\setlength{\captionmargin}{10.0pt}	
	\centering
	\includegraphics[width=\columnwidth]{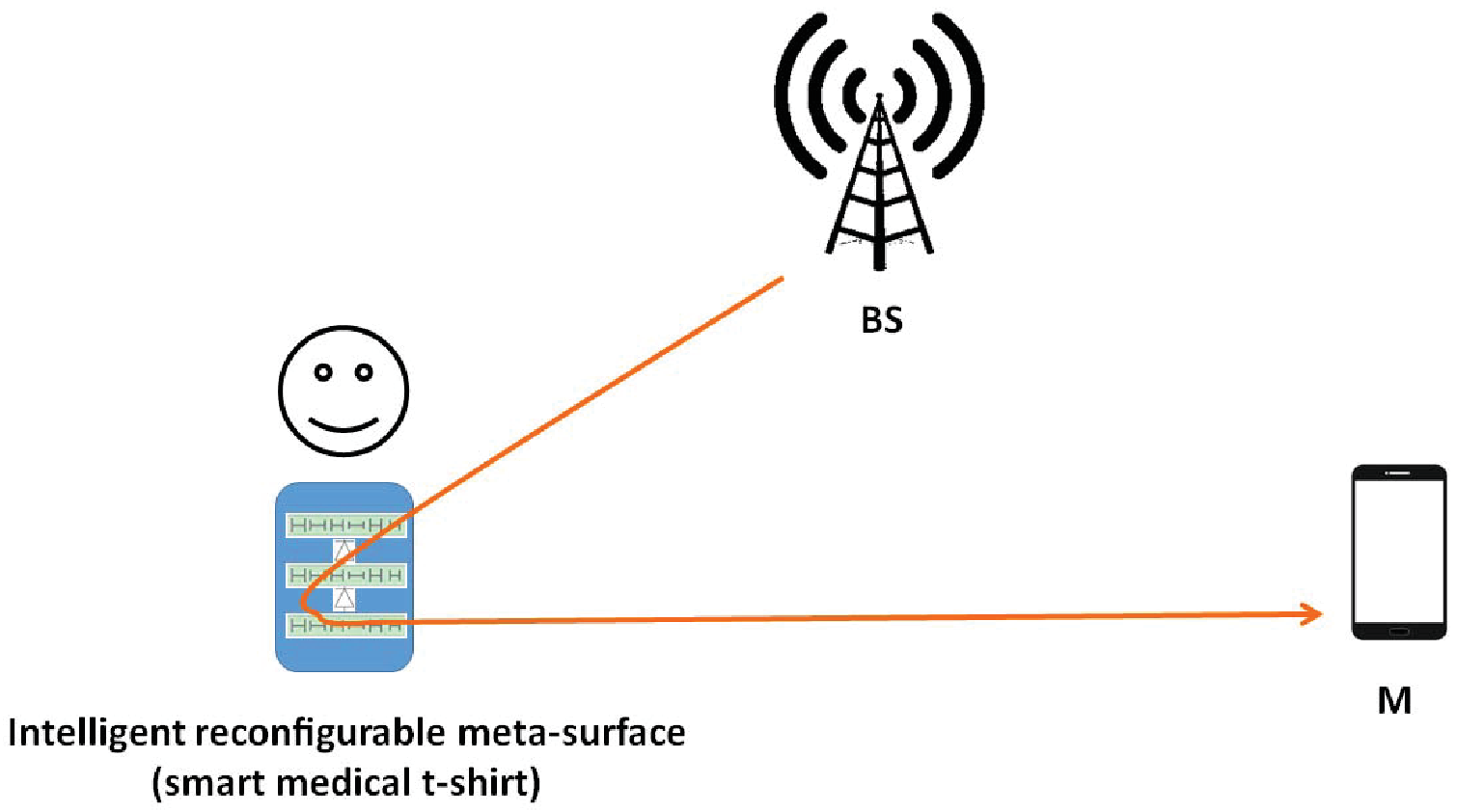}
	\caption{Smart radio environments: Sensing and computing}
\end{figure}
\begin{figure}[!t] \label{MetasurfaceModulation}
	\setlength{\captionmargin}{10.0pt}	
	\centering
	\includegraphics[width=\columnwidth]{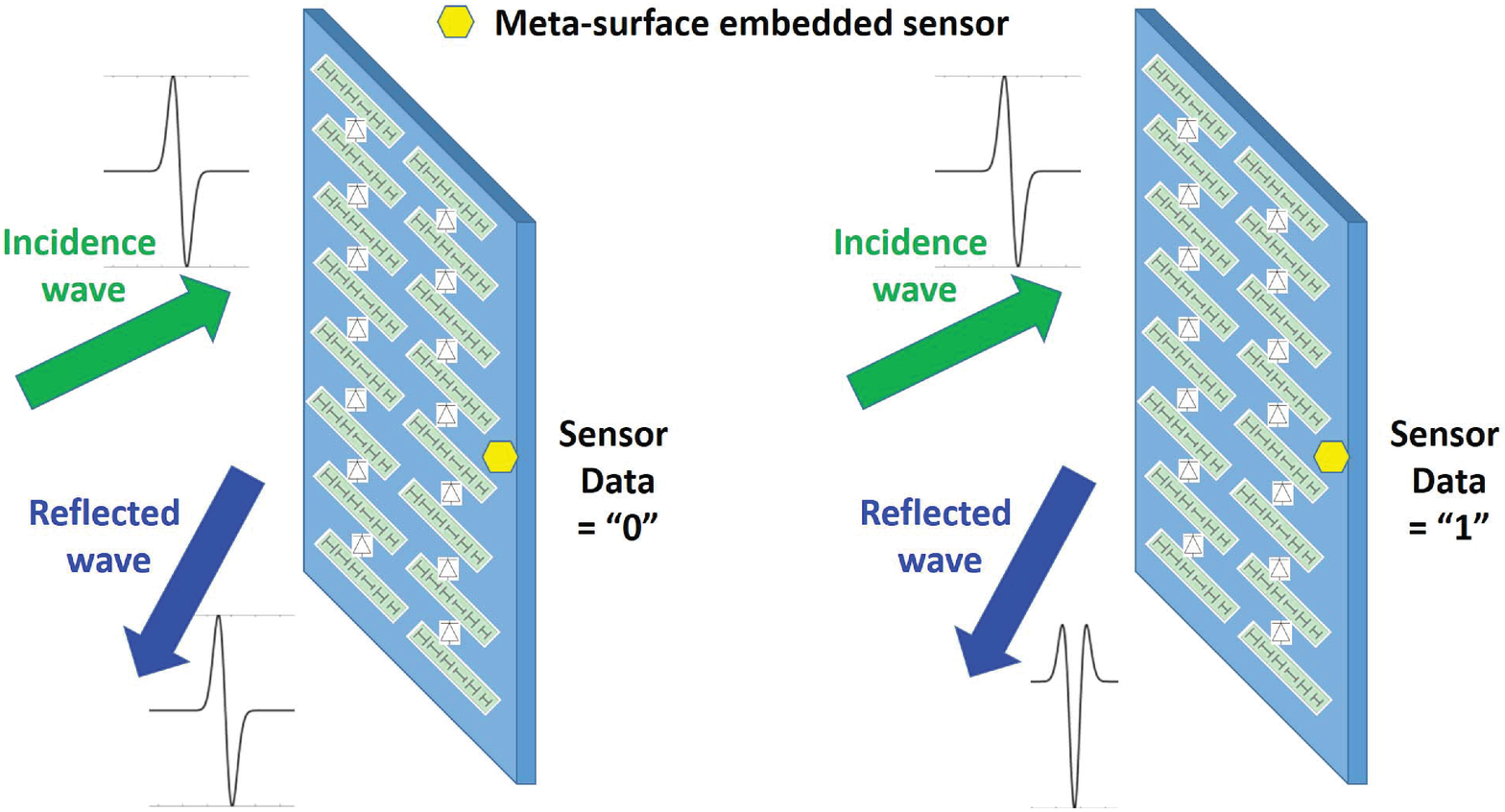}
	\caption{Working principle of meta-surface based modulation}
\end{figure}
\subsection{Reconfigurable Environments: Sensing and Computing}
Let us now consider Fig. 4, which shows a person wearing a smart t-shirt with embedded sensors for health monitoring. In a conventional wireless system, each time the sensor needs to report the sensed data to a mobile terminal M, it has to emit a new radio wave that drains its small size battery \cite{2} and increases the spectrum usage. This is overcome if the t-shirt is coated with an intelligent reconfigurable meta-surface. Rather than emitting a new signal, the data of the sensor can be embedded into the reflected signal from the BS. Assume, for simplicity, that the sensor needs to report a single bit of information. If the bit is ``0'', the intelligent reconfigurable meta-surface does not alter the reflected signal from BS1. If the bit is ``1'', the intelligent reconfigurable meta-surface encodes the bit into the reflected wave from the BS, by, e.g., differentiating it. This is depicted in Fig. 5. Generally speaking, the intelligent reconfigurable meta-surfaces can introduce a delay by first storing and then releasing the reflected signal \cite{18}, or can modify the emitted waveform by differentiating/integrating the reflection \cite{IL_Computing}, \cite{19}, or by encoding the bit into a specified value of its reflection coefficient. This approach allows the sensor to report the data to the mobile terminal M without emitting any signals and in an energy-free manner. It simply recycles the reflected signal that originates from the BS. This approach, in particular, allows battery-constrained tiny devices to take advantage of the signals emitted by large-size devices with less stringent energy constraints, and to convey their own sensed data for free. This approach, which we will call meta-surface based modulation, is new and can be considered a major generalization of distributed spatial modulation that was introduced in \cite{24}. Thanks to this new concept, wireless network operators can offer new services without emitting additional radio waves, and without adding batteries into the environment. They simply recycle their own already existing radio waves. It is worth mentioning that the concept of meta-surface based modulation can be applied to the network scenario depicted in Fig. 3 as well, in order to enable long-range transmissions. These applications are discussed in the sequel. \\ \\

\subsection{A New Communication-Theoretic Model}
The case studies illustrated and discussed in the previous two sections highlight that, in current wireless networks, the devices and transmission protocols are usually designed and optimized to adapt themselves to the radio environment. Smart radio environments are fundamentally different: Rather than optimizing (only) the endpoints, i.e., the devices, the radio environment is dynamically configured and assists the transfer and processing of information between the devices. Potentially, the endpoint radios can be made as simple as possible, with major economic advantages for wireless network operators \cite{21}. \\  \\

Broadly speaking, we can say that current wireless networks operate according to three main postulates: \\
\begin{enumerate}
   \item The environment is usually perceived as an ``unintentional adversary'' to communication and information processing. \\
   \item Only the end-points of the communication network are usually optimized. \\
   \item Wireless network operators have usually no control of the environment. \\ \\
 \end{enumerate}

Smart radio environments, on the other hand, provide wireless network operators with new degrees of freedom to further improve the network performance, since the environment is not viewed as a passive entity and it not taken for granted, but can be customized as the wireless network operators desire.  \\ \\

\begin{figure}[!t] \label{CommTheoryModel}
	\setlength{\captionmargin}{10.0pt}	
	\centering
	\includegraphics[width=\columnwidth]{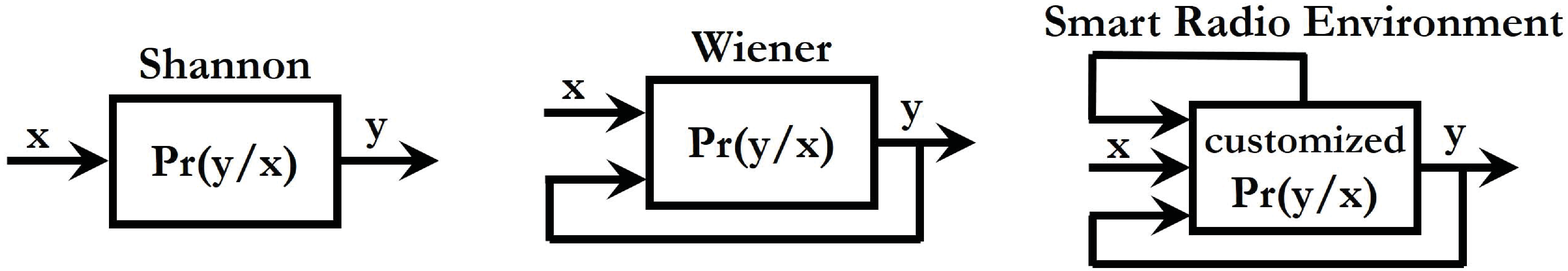}
	\caption{New communication-theoretic model for smart radio environments}
\end{figure}
Conceptually, the difference between current wireless networks and smart radio environments is depicted in Fig. 6. According to Shannon \cite{25}, the system model is given and is formulated in terms of transition probabilities (i.e., $\Pr ( {{y \mathord{\left/{\vphantom {y x}} \right. \kern-\nulldelimiterspace} x}} )$). According to Wiener \cite{26}, the system model is still given, but its output is fed back to the input, which is optimized by taking the output into account. For example, the channel state is sent from a receiver back to a transmitter for channel-aware beamforming. In smart radio environments, by contrast, the environmental objects are coated with meta-surfaces that are capable of sensing the system's response to the radio waves (the physical world), and of feeding this response back to the input (the digital world or network controller) \cite{3}. Based on the sensed data, the input and the wave manipulations applied by the meta-surfaces can be jointly optimized and configured through a software controller. For example, the input signal can be steered towards a meta-surface, which is configured to reflect it towards a given receiver, which is, in turn, steered towards the meta-surface. \\

In contrast to Shannon's and Wiener's models, which have been widely researched during the last decades, the theoretic and algorithmic foundation of the system model for smart radio environments is unknown. In the next sections, we will elaborate on the fundamental gaps of knowledge that need to be addressed towards filling this fundamental open research issue. A fundamental question, in particular, immediately raises to our attention by directly inspecting Fig. 6: \\

\textit{To be optimally configured, how much sensed and feedback data do smart radio environments need}? \\ \\

From Fig. 6, in fact, it is apparent that, to turn the concept of smart radio environments into a reality, a critical issue to address is constituted by the amount of sensed data that the meta-surfaces need to gather and to make available (feedback) to an overarching network controller in order to be able to configure and optimize the environment as a function of the network conditions. Efficient solutions need to be developed in order to reduce the amount of sensed data for network optimization, and, at the same time, in order to make it available (i.e., report, transmit) to the network controller with low overhead and high energy efficiency. \\ \\

\subsection{Novelty Compared with Current Wireless Networks}
In this section, we briefly compare smart radio environments against widely employed technologies to enhance the performance of wireless networks. To better elucidate the difference and significance of smart radio environments, we consider, as an example, a typical cellular network. \\ \\

\textbf{Current wireless networks} – The distinguishable feature of cellular networks lies in the users' mobility. The locations of the base stations cannot, in general, be modified according to the users' locations. Some exceptions, however, exist \cite{27}, \cite{28}, and we elaborate on them below. The mobility of the users throughout a location-static deployment of base stations renders the user distribution uneven throughout the network, which results in some base stations to be overloaded and some others to be underutilized. This is a known issue in cellular networks, and is tackled in different ways. \\

Two interlinked approaches are load balancing \cite{29} and the densification of base stations. Network densification is a promising solution, but it has its own limitations \cite{30}. It is known, e.g., that network densification increases the network power consumption as the number of base stations per square kilometer increases. This is exacerbated even more with the advent of the Internet of Things (IoT), where the circuit power consumption increases with the number of users per square kilometer \cite{143}. Ultra-dense network deployments, also, enhance the level of interference, which needs to be appropriately controlled in order to achieve good performance \cite{30}. In addition, each base station necessitates a backhaul connection, which may not always be available. \\

Other solutions based on Massive Multiple-Input-Multiple-Output (MIMO) schemes could be employed, but they usually necessitate a large number of individually controllable radio transmitters and advanced signal processing algorithms \cite{31}. Similar comments (i.e., power consumption, hardware complexity, blocking of links, etc.) apply to using millimeter-wave communications \cite{32}, \cite{33}. It is worth mentioning that millimeter-wave systems can take advantage of reconfigurable meta-surfaces as a source of controllable reflectors that can overcome non-line-of-sight propagation conditions, and can enable the otherwise impossible communication among the devices \cite{10}. Reconfigurable intelligent meta-surfaces that act as reconfigurable reflectors, in particular, constitute a promising solution to establish strong non-line-of-sight links whenever the line-of-sight is not available or it is just not sufficiently strong to achieve a good connectivity or a high throughput. This is often the case of signal transmission in high frequency bands, which include millimeter-wave and beyond 100 GHz communications. \\

Without pretending to be exhaustive, other relevant solutions that are typically used in wireless encompass retransmission methods that negatively impact the network spectral efficiency, the deployment of specific network elements, e.g., relays, which increase the network power consumption as they are made of active elements (e.g., power amplifiers), and that either reduce the achievable link rate if they operate in half-duplex mode or are subject to severe self-interference if they operate in full-duplex mode \cite{MDR-Relay-1}, \cite{MDR-Relay-2}. It is worth mentioning, in particular, that intelligent reconfigurable meta-surfaces that act as reconfigurable reflectors are different from relays, since their main functionality is to reconfigure the multi paths in a way that they are optimally combined at the intended destination. In addition, they are not affected by the self-interference and by the noise amplification effects, since reflectors are not affected by such impairments. \\ \\

\textbf{Smart Radio Environments} – In contrast to the aforementioned technologies, smart radio environments are fundamentally different. The reconfigurable meta-surfaces can be made of low-cost passive elements that do not require any active power sources for transmission \cite{34}. Their circuitry and embedded sensors can be powered with energy harvesting modules as well \cite{72}, \cite{35}: An approach that has the potential of making them truly energy-neutral. They do not apply any sophisticated signal processing algorithms (coding, decoding, etc.), but primarily rely on the programmability and re-configurability of the meta-surfaces, and on their capability of appropriately shaping the radio waves impinging upon them \cite{36}. They can operate in full-duplex mode without significant or any self-interference, they do not increase the noise level, and do not need any backhaul connections to operate. Even more importantly, the meta-surfaces are deployed where the issue naturally arises: Where the environmental objects, which, in current wireless networks, reflect, refract, distort, etc. the radio waves in undesirable and uncontrollable ways, are located. The cost, however, comes from the overhead that is needed for controlling the intelligent reconfigurable meta-surfaces. \\

\begin{figure}[!t] \label{Routing}
	\setlength{\captionmargin}{10.0pt}	
	\centering
	\includegraphics[width=\columnwidth]{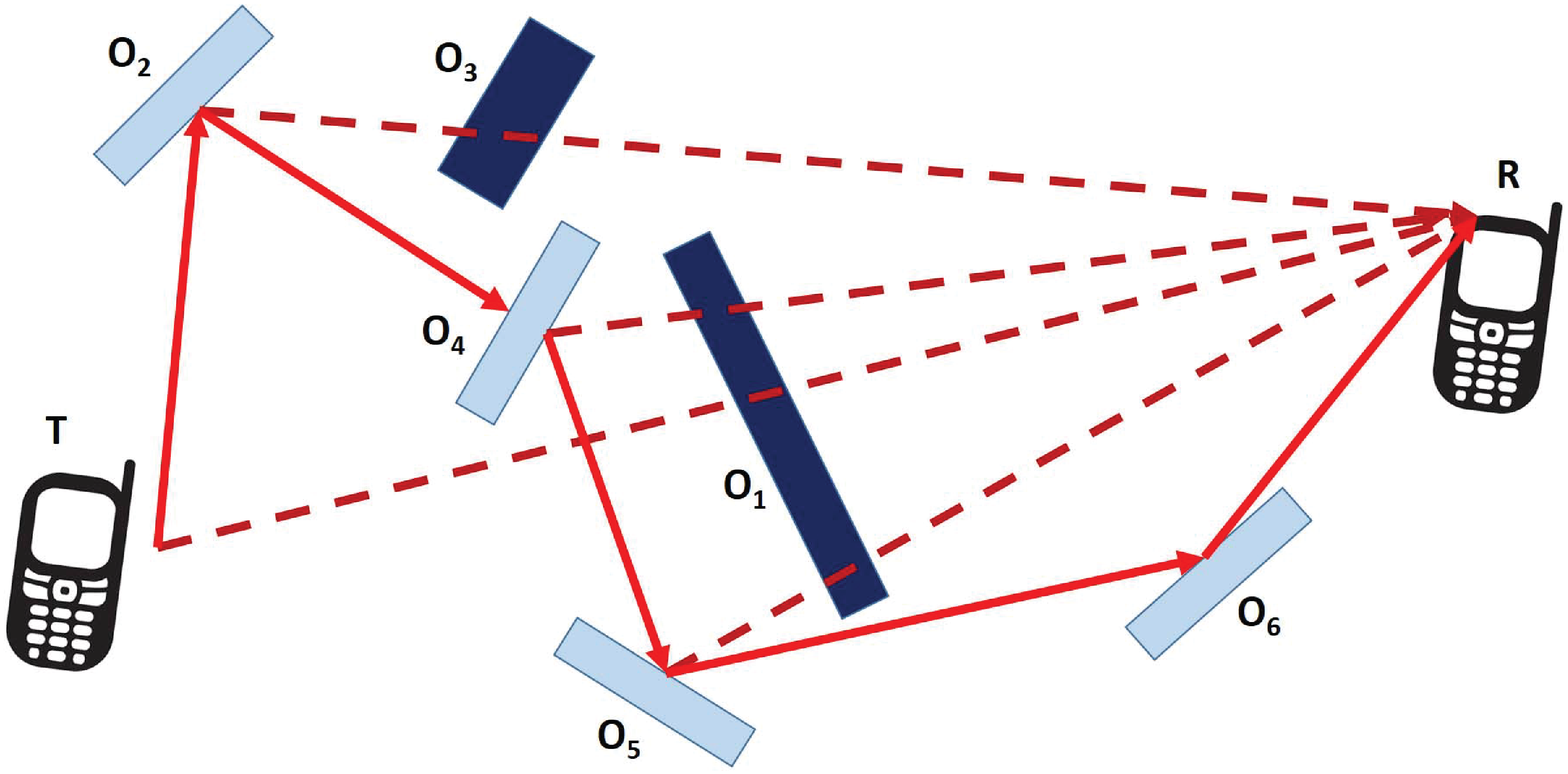}
	\caption{Smart ratio environments: Environmental routing}
\end{figure}
Since the input-output response of the meta-surfaces is not subject to conventional Snell's laws anymore \cite{15}, the locations of the objects that assist a pair of transmitter and receiver to communicate, and the functions that they apply to the received signals can be chosen to minimize the impact of multi-hop-like signal attenuation, as illustrated in Fig. 7. In addition, the phases of the scattering particles that constitute the meta-surfaces can be optimized to coherently focus the waves towards the intended destination without using active elements. These functionalities, in addition, are transparent to the base stations and the mobile terminals, as there is no need to change their hardware and software. \\

The specific characteristics and properties of the meta-surfaces can be exploited to recycle existing radio waves and to foster the seamless integration of communications with sensing, storage, and computing (see Fig. 5). Furthermore, the number of environmental objects can potentially exceed the number of antennas at the endpoint radios, which implies that the available options for system optimization can potentially exceed that of current wireless network deployments \cite{6}. The freedom of controlling the response of each meta-surface and choosing its location via a software-programmable interface makes, in addition, the optimization of wireless networks agnostic to the underlying physics of wireless propagation and the meta-materials. It enables, in addition, the seamless integration of reconfigurable meta-surfaces into software networks. Further information about the programmability via software and the integration of reconfigurable meta-surfaces into software networks can be found in \cite{VISOSURF-WEB}. \\

Finally, despite the practical challenges of deploying robotic (terrestrial) base stations that are capable of autonomously moving throughout a given region \cite{27}, \cite{28}, experimental results conducted in an airport environment, where the base stations were deployed on a rail located in the ceiling of a terminal building \cite{37}, showed promising gains. The possibility to deploy mobile reconfigurable meta-surfaces is, on the contrary, practically viable. The meta-surfaces can be easily attached to and removed from objects (e.g., facades of buildings, indoor walls and ceilings, advertising displays), respectively, thus yielding high flexibility for their deployment. The position of small-size meta-surfaces on large-size objects, e.g., walls, can be adaptively optimized as an additional degree of freedom for system optimization: Thanks to their 2D structure, the meta-surfaces can be mechanically displaced, e.g., along a discrete set of possible locations (moving grid) on a given wall. \\ \\

\section{Communication-Theoretic and Algorithmic Foundation}
In this section, we summarize current research activities that are related to the concept of smart radio environments. It is apparent that the concept of reconfiguring the radio environment was implicitly proposed by a few authors, but only recently it received major attention from the research community. \\ \\

\subsection{Current Research Landscape}
The idea of reconfiguring the radio environment to make it smart, in contrast to current wireless networks, has emerged only recently and in different forms. Notable examples include: Intelligent walls \cite{7}, \cite{8}, smart reflect-arrays \cite{9}-\cite{11}, and low-cost devices embedded into walls \cite{6}, which are viewed as a contiguous surface of electromagnetic material \cite{13}, \cite{38}-\cite{Slim}. The solution analyzed in \cite{13} and \cite{38}, in particular, is based, in contrast to other research efforts, on using electromagnetically active material, which entails an increase of the power consumption. In our view of smart radio environments, the intelligent reconfigurable meta-surfaces are, on the other hand, made of passive or almost passive electromagnetic material. The authors of \cite{39}-\cite{Slim}, in particular, focus their attention on the performance evaluation and optimization of passive intelligent reconfigurable meta-surfaces, which act as tunable and reconfigurable reflectors. A special instance of smart radio environments is media-based modulation \cite{44}-\cite{46}, in which reconfigurable Radio Frequency (RF) mirrors, which act as on-off switches, are deployed around a radiating element and are used to encode information onto the different channel states obtained by configuring the status of the switches. Media-based modulation is a single-RF modulation scheme, similar to beamspace MIMO \cite{47}, spatial modulation \cite{48}, \cite{49}, load modulation \cite{50}, and index modulation \cite{51}. The use of reconfigurable meta-surfaces to make the environment reconfigurable is currently being researched under the auspices of the European-funded VISORSURF project \cite{22}. The vision of the project consists of coating the environmental objects with reconfigurable meta-surfaces whose response to the radio waves is programmed in software \cite{3}. Compared with other solutions, the use of meta-surfaces have major economic benefits, e.g., they reduce the waste of resources \cite{21}, and offer a more accurate control of the radio waves, and a better deployment scalability \cite[Table 1]{20}. \\

The use of meta-surfaces to make the radio environment smart relies on three technological breakthroughs: 1) the possibility of fabricating meta-surfaces with arbitrary and controllable wave manipulation functionalities, 2) the possibility of making the response and functionality of the meta-surfaces reconfigurable based on the network conditions, and 3) the possibility of controlling the meta-surfaces via a software-defined interface that enables their seamless integration into the software-defined networking paradigm \cite{3}, \cite{4}. Luckily, these technology enablers are today possible with current technology. \\

Meta-surfaces that synthesize several types of wave transformation across the entire telecommunication spectrum can be efficiently fabricated, \cite{IL_1}-\cite{IL_END}, \cite{52}-\cite{63}. For example, methods for the synthesis of meta-surfaces with complete control of the transmitted and reflected waves \cite{64}, and efficient computational frameworks to optimize large-area meta-surfaces \cite{65} are available. The meta-surfaces are suited for applications beyond communications, e.g., sensing signals \cite{17}, \cite{66}, storing and releasing data \cite{18}, \cite{67}, and analog computing \cite{IL_Computing} \cite{19}, \cite{68}, \cite{69}. This is instrumental in order to exploit them for sensing besides communications, and in order to interconnect the physical and digital worlds in a seamless manner, as described in the previous sections and illustrated in Fig. 4 and Fig. 5. \\

The reconfigurability of the meta-surfaces is currently possible in different ways \cite{36}. In \cite{22}, e.g., a network of miniaturized controllers, passive patches, and active binary switches is embedded throughout the meta-surface. The status of the switches determines the response of the meta-surface. To make the meta-surfaces as passive as possible, in addition, energy harvesting modules are used. The intelligent reconfigurable meta-surfaces, in fact, are expected to consume much less power than current devices, and, thus, energy harvesting solutions are considered to be a viable option. In \cite{22}, the controllability in software of the meta-surfaces is ensured via a gateway that connects the meta-surface to an external controller: Based on the sensed data, the controller computes the best wave transformations to apply, and sends the configuration of the switches that synthesizes them to the meta-surface. This operating principle fulfills the communication-theoretic model in Fig. 6, and enables the integration of the reconfigurable meta-surfaces into the software-defined networking paradigm \cite{4}, \cite{70}. Protocols to enable the exchange of data within the reconfigurable meta-surfaces with stringent energy, latency, and robustness are available as well \cite{71}-\cite{76}. In these papers, it is proved, in particular, that wireless technologies in the millimeter and tera-hertz bands are suitable to realize nano-networking protocols for enabling the transmission of data within the meta-surface. \\ \\

\subsection{Fundamental Gaps of Knowledge: A Communication-Theoretic Perspective}
From the state-of-the-art assessment elaborated in the previous section, it is apparent that smart radio environments enabled by intelligent reconfigurable meta-surfaces are a recent but feasible technology. More precisely, different types of meta-surfaces can be efficiently fabricated, can be made reconfigurable, and can be controlled in software. Current research activities are, however, focused on implementing testbeds, fabricating new meta-surfaces, and designing nano-networking and software-defined protocols. These research efforts are primarily related to the research fields of physics with focus on meta-materials and electromagnetism, computer science with focus on software-defined networking, and on system-on-chip design with focus on nano-communications and networking protocols. These aspects are, in particular, well tackled and discussed in several recent papers published by the scientists of the VISORSURF project \cite{VISOSURF-WEB}. \\ \\

In the present paper, on the other hand, we focus our attention on the fundamental gaps of knowledge towards realizing the vision of smart radio environments from the point of views of communication theory and wireless communications. Based on the state-of-the-art assessment elaborated in the previous section, in fact, the theoretic and algorithmic foundation of smart radio environments are unexplored in the context of communication theory and wireless communication research. More precisely, the following fundamental research questions have no answers yet: \\

\begin{enumerate}
  \item \textit{What are the fundamental performance limits (overhead included) of large-scale wireless networks in the presence of intelligent reconfigurable meta-surfaces}? \\

  \item \textit{What analytical methodologies to use for unveiling such fundamental performance limits}? \\

  \item \textit{What algorithms and protocols to use for achieving such fundamental performance limits}? \\

  \item \textit{What is the optimality of such algorithms and protocols, and are they implementable in practice}? \\

  \item \textit{What algorithms and protocols to use in order to leverage smart radios environments as a unified platform that integrates communications, sensing, and computing}? \\

  \item \textit{What simulation tools to use for validating the predicted ultimate performance limits and scaling laws, as well as to evaluate the achievable performance of the proposed algorithms and protocols in realistic large-scale wireless networks}? \\

    \item \textit{What is the economic impact of smart radio environments on the upcoming 5G and future 6G markets, and is it sustainable}? \\

    \item \textit{What are the performance gains of smart radio environments compared with current wireless networks}? \\ \\
\end{enumerate}

From the communication-theoretic standpoint, smart radio environments can be viewed as the synergistic amalgamation of intelligent reconfigurable meta-surfaces and large-scale wireless networks. With this in mind, we have identified four fundamental and open research issues that need to be solved in order to answer the above mentioned questions. In the following paragraphs, we will elaborate on these four fundamental gaps of knowledge in communication theory and wireless communications, and we will briefly discuss promising approaches to tackle them. We consider them, in fact, fundamental components to lay the communication-theoretic and algorithmic foundation of smart radio environments, and therefore to enable the seamless integration of intelligent reconfigurable meta-surfaces in large-scale wireless networks. \\ \\

\textbf{A. The Need of Computational Analytical Frameworks for the Synthesis and Analysis of Reconfigurable Meta-Surfaces in Smart Radio Environments - The Role of the Zero-Thickness Sheet Model} \\

\textit{How to incorporate the physical structure and characteristics of many spatially distributed meta-surfaces (of generic geometry and shape) into utility functions, beyond the electromagnetic field, that are relevant to design wireless networks that are deployed over large geographical areas}? \\

The answer to this question is the essence of the first fundamental gap of knowledge that we are faced with. To the best of our knowledge, there exists no analytical approach that allows us to account for the wave manipulations introduced by reconfigurable meta-surfaces into utility metrics that are of interest in communication theory, e.g., the coverage probability, the spectral efficiency, the energy efficiency, the delay, etc. \\

A suitable approach to overcome this issue consists of using the so-called zero-thickness sheet model for the meta-surfaces, which can be used for planar and conformal meta-surfaces \cite{77}, \cite{78}, \cite{84}. Based on this model, the meta-surfaces are assumed to have zero-thickness, and their geometric and electromagnetic parameters are mapped onto specified waves' transformations by using surface susceptibility tensors. By using this approach, the meta-surfaces can be modeled as systems of arbitrary input-output response, which can be optimized to maximize some given utility functions. The main issue is that this modeling approach is efficient to formulate the discontinuity of the electromagnetic field of individual meta-surfaces, but closed-form solutions are not always available \cite{80}. How to use this model for system-level performance analysis and optimization is, in addition, unknown at present, and it requires one to commence the analysis directly from Maxwell's equations. \\

Overcoming this gap of knowledge on how to amalgamate communication-theoretic models for wireless signals and wireless networks with electromagnetic-theoretic models for reconfigurable meta-surfaces is a necessity for analyzing and optimizing smart radio environments. To elucidate the significance of this problem, let us consider an example. Suppose that one succeeds in obtaining an expression of the spectral efficiency of a cellular network as a function of the surface susceptibility tensors of the meta-surfaces that coat the walls of buildings, and that one succeeds in identifying the best surface susceptibility functions that optimize the spectral efficiency. This achievement will be instrumental for two reasons \cite{77}: 1) if the susceptibility functions can be implemented in practice, i.e., there exist physical structures that synthesize them, then we will be able to optimize the wave manipulations of the meta-surfaces in large-scale networks; and 2) if the susceptibility functions cannot be implemented in practice, i.e., there exist no physical structures that synthesize them, then we will be capable of identifying practical meta-surfaces that yield a close-to-optimal spectral efficiency, will be able to quantify the loss with respect to the optimum, and will be able to unveil the constraints to impose on communications-related parameters, e.g., the density and transmit power of the base stations in a cellular networks or the density of meta-surfaces, to obtain optimal performance but with surface susceptibility tensors that can be realized. \\

This simple example clearly illustrates the necessity and relevance of developing computational analytical frameworks for the synthesis and analysis of reconfigurable meta-surfaces in smart radio environments. \\ \\

\textbf{B. Tractable Analytical Frameworks for Modeling, Analyzing, and Optimizing Smart Radio Environments in Large-Scale Wireless Networks - The Role of Random Spatial Processes} \\

The integration of reconfigurable meta-surfaces into a wireless network is not only limited to identifying an electromagnetic-based  and analytically tractable physical model for the meta-surfaces. The meta-surfaces are expected to be attached to environmental objects or be even part of the fabrics of the objects themselves, e.g., the facades of buildings, the walls of rooms, etc. The environmental objects are, in particular, distributed in space according to very complex spatial patterns. Besides the need of parametric, computational, and electromagnetic-compliant models for the meta-surfaces and the need of incorporating them into the signal models used in wireless, the modeling and optimization of smart radio environments necessitate spatial models that account for i) the distribution of the locations of the meta-surfaces in large-scale wireless networks, ii) the wave manipulations applied by the meta-surfaces depending on their spatial locations and on the radio waves impinging upon them, and iii) the spatial locations and wave manipulations applied by other randomly distributed meta-surfaces. We lack these tractable models in communication theory and wireless networks. \\

Let us elaborate a little bit further on the reasons behind this fundamental gap of knowledge. The most suitable analytical tools and spatial models for representing the locations of the transmitters, receivers, and environmental objects in large-scale wireless networks are stochastic geometry and random spatial processes \cite{85}-\cite{100}. In 2011, notably, Poisson point processes were used to formulate the coverage probability in cellular networks \cite{101}. Since then, their application to the modeling and analysis of wireless networks has been relentless \cite{102}-\cite{104}. Also, spatial processes have found many applications beyond communications, e.g., localization \cite{105}-\cite{107}, caching (local data storage) \cite{108}-\cite{110}, distributed sensing and data fusion \cite{111}-\cite{113}. Fundamental issues, however, remain open to use them for modeling, analyzing, optimizing wireless networks \cite{114}. When it comes to modeling the spatial distribution of intelligent reconfigurable meta-surfaces and to incorporate the wave manipulations applied by the meta-surfaces, in particular, it is not difficult to realize that the models applied to date cannot be applied, since they are based on assumptions that are not compliant with the operations of the meta-surfaces. \\

In order to understand these fundamental limitations of current models, let us consider a concrete example. In smart radio environments, the environmental objects are coated with intelligent reconfigurable meta-surfaces that can reflect, refract, absorb, and modulate data onto the received signals. In the current literature, the environmental objects are always modeled as entities that can only attenuate the signals, by making the links either line-of-sight or non-line-of-sight \cite{116}-\cite{121}. Modeling anything else is known to be difficult. In \cite{122}, the authors have investigated the impact of reflections, but only based on conventional Snell's laws. This work highlights the analytical complexity, the relevance, and the non-trivial performance trade-offs: The authors emphasize that the trends highly depend on the fact that the total distance of the reflected paths is almost always two times larger than the distance of the direct paths. This occurs because, based on Snell's law, the angles of incidence and reflection are the same. \\

\textit{What if meta-surfaces-coated randomly distributed objects optimize the reflected signals in directions possibly different from those predicted by the Snell’s law}? \textit{What if the signals’ propagation is altered in ways different than reflection}? \\

Generally speaking, none of the currently available tools can be applied because the way how radio waves are reflected depend on the position of the transmitters and receivers, which has never been the case before. These two questions exemplify the fundamental gap of knowledge that we are faced with. Only recently, we have introduced in \cite{115} the first approach that allows one to compute the probability that a randomly distributed meta-surface can act as a reflector under the assumption that it can reflect signals originating from any possible directions and towards any possible directions. The physical response of the meta-surfaces is, however, not taken into account. \\

Besides modeling the impact of environmental objects and meta-surfaces as blocking elements, several other fundamental modeling issues need to be overcome to be able to incorporate the intelligent reconfigurable meta-surfaces into wireless networks. This includes the impact of spatial correlations among randomly distributed meta-surfaces, e.g., \cite{123}-\cite{136}, \cite{155}, the impact of near-field propagation effects that need to be taken into account if large-scale meta-surfaces are deployed in confined environments, e.g., in indoor settings, and the compelling need of developing abstraction models that are suitable and amenable for system optimization \cite{146}-\cite{148}. \\ \\

\textbf{C. Design of Communication Protocols for Seamlessly Integrating Communications, Sensing, and Computing - The Role of Spatial Modulation} \\

As mentioned in previous sections, future wireless networks will not only offer communication services but will be an integrated platform that is intended to provide the users with communications, sensing, computing (or, more in general, distributed information processing), and localization services by using the same network infrastructure. \\ \\

Then, two fundamental questions naturally arise in the context of smart radio environments: \\

\begin{enumerate}
  \item \textit{Are intelligent reconfigurable meta-surfaces a suitable technology to realize such a platform that integrates communications and distributed information processing}? \\
  \item \textit{If so, what algorithms and protocols to use in order to seamlessly interconnect the physical and digital world by using intelligent reconfigurable meta-surfaces}? \\ \\
\end{enumerate}

We are not aware of any solutions and proposals to address this fundamental gap of knowledge. In the previous sections, in addition, we have emphasized that a fundamental difference between smart radio environments and current network models (see our proposed communication-theoretic model in Fig. 6), lies in the amount of sensed data and feedback overhead that is needed in order to optimize and customize the environments, i.e., to optimize the wave transformations applied by the reconfigurable meta-surfaces. While the reconfigurable meta-surfaces offer a unique platform for distributed sensing, distributed computing, and distributed information processing, two major issues deserve attention: \\

\begin{enumerate}
  \item The amount of sensed data that is necessary for optimizing the reconfigurable meta-surfaces needs to be reduced as much as possible. \\
  \item The sensed data needs to be reported at low energy, power, and bandwidth cost, in order to avoid to be the bottleneck of the overall system. \\ \\
\end{enumerate}

We commence by elaborating the second issue and postpone the discussion of the first issue to further text below. The fundamental reason why sensing is, in general, not resource efficient lies in the fact that, in wireless communications, the devices generate new signals every time that they have to transmit data. In other words, the sensors consume power and bandwidth, as well as increase the level of interference and the usage of spectrum, every time that they need to report their data. This needs to be avoided. In Fig. 5, we have described how this issue can be overcome by employing the principle of meta-surface based modulation, which foresees to encode the data sensed by the meta-surfaces into specified wave transformations of the meta-surfaces themselves. By using this approach, the sensed data is piggybacked into the signals received by the meta-surfaces and emitted by other devices, thus providing an efficient solution for transmitting data without generating new signals. For example, signals' reflections are employed for encoding data in a resource-free manner. It is worth mentioning that with the term ``sensed data'' we refer to the data that is sensed to optimize the wave transformations of the intelligent reconfigurable meta-surfaces, and to the data sensed for other different purposes, e.g., health monitoring (Fig. 4). \\

Our proposed approach has some similarities, but major differences, with bistatic backscatter communications \cite{157}. Despite the recent research activities in backscatter communications, in fact, major limitations in terms of tradeoff among data rate, error rate, communication range, and energy efficiency exist \cite{158}. Similar to bistatic backscatter communications, meta-surface based modulation consists of modulating the data of sensors available in the meta-surfaces for various applications, into the signals, e.g., reflected or refracted, by the meta-surfaces and that originate from other transmitters. This approach results in a distributed sensing platform that does not need any energy for transmitting the sensed data because the signals emitted by other devices for other purposes are used instead. The sensed data can be modulated, e.g., onto the reflection coefficient or the radiation pattern of reconfigurable meta-surfaces. \\

Besides being a promising enabler for realizing a distributed platform for interconnecting the digital and physical worlds in a seamless manner, meta-surface based modulation is also a promising solution to report feedback data and efficiently collect the necessary contextual information to optimize the operation of reconfigurable meta-surfaces. More precisely, the feedback data can be embedded, e.g., onto the reflections of other signals without necessitating any additional resources. Recent results, in fact, show that data can indeed be encoded into reconfigurable features of radiating elements, such as antennas \cite{144}, \cite{159}, \cite{160}. This motivates the principle of meta-surface based modulation as an enabling transmission protocol to report feedback data without using extra resources. The theoretic limits and practical algorithms to leverage this approach are, however, unknown. \\ \\

\textbf{D. System-Level Simulation of Large-Scale Wireless Networks in the Presence of Reconfigurable Meta-Surfaces} \\

\textit{How to efficiently simulate a large-scale wireless network, where each environmental object is coated with a reconfigurable meta-surface whose waves' transformations adhere to the generalized Snell's laws, and are chosen to maximize network-wide utility functions that account for the physical structure, the finite size, and geometry of each meta-surface}? \\

This question constitutes a major gap of knowledge for analyzing and optimizing smart radio environments. To the best of our knowledge, in fact, there exist no simulators that account for general (in agreement with the generalized Snell's laws) wave transformations that can be realized by reconfigurable meta-surfaces randomly distributed in large-scale wireless networks \cite{175}. \\ \\

The fundamental reasons at the origin of the lack of such simulators are the following: \\

\begin{itemize}
  \item A meta-surface is a highly complex structure: It is electrically thin, is electrically large, and is made of sub-wavelength particles. Because of these peculiar characteristics, no commercial software is capable of efficiently simulating meta-surfaces that are modeled as a sheet of zero thickness \cite{79}-\cite{81}. Efficient numerical algorithms that generalize finite difference time or frequency domain methods and account for individual meta-surfaces have been recently proposed in \cite{79}, \cite{84}. However, they are not scalable for application in large-scale wireless networks. \\
  \item Ray optics modules available in commercial multi-physics simulators implement conventional Snell's laws \cite{3}. In \cite{20}, ad hoc rotations of the spatial derivatives of the meta-surfaces are applied to overcome this limitation. The approach, however, is applicable only to planar meta-surfaces, is an approximation, and is difficult to generalize.  \\
  \item Due to memory and computation time, it is not possible to simulate an entire (large-scale) wireless network by using a full-wave simulator that models the meta-surfaces as a zero-thickness sheet \cite{80}, \cite{170}. \\ \\
\end{itemize}

It is necessary, therefore, to develop system-level simulators that integrate ray optics modules that are in agreement with generalized Snell's laws, and that, more in general, allow us to account for general wave transformations that can be applied by the reconfigurable meta-surfaces. The availability of such system-level simulators is essential in order to substantiate new theoretical models and scaling laws, as well as to test and optimize new algorithms and protocols in realistic environments. Some interesting preliminary results based on a graph-based model for the smart radio environments can be found in \cite{RECENT}. \\ \\

\textbf{E. Environmental AI: AI for Smart Radio Environments} \\

It is apparent that smart radio environments are a very complex system to design. This originates from the large number of parameters to be optimized based on the contextual information that is gathered by the intelligent reconfigurable meta-surfaces and that is made available to the network controller. As depicted in Fig. 6, this usually requires a large amount of sensed data from the sensors embedded into the reconfigurable meta-surfaces. Collecting, processing, and reporting this large amount of data are usually resource consuming, since these operations need to be executed every time that the network conditions change, e.g., the channel changes, the positions of the users change, etc. Therefore, as already mentioned, it is extremely important to reduce the amount of sensed data that is necessary for optimizing the operation of the intelligent reconfigurable meta-surfaces. \\

In an era where machine learning is considered to be a pervasive and effective solution for addressing several complex problems, it is legitimate to investigate its role in the context of smart radio environments \cite{150}, \cite{152}. This is especially true, in particular, in light of the recently approved ``ITU-T Y.3172 architectural framework for machine learning in future networks including IMT-2020'' \cite{ML_ITU}. \\

In principle, machine learning methods are powerful approaches for optimizing the reconfigurable meta-surfaces. Reinforcement learning, in particular, implements the learning and decision-making procedures by interacting with the environment: Taking actions and receiving feedback on the result of the actions that are taken. By using this approach, we can envision reconfigurable meta-surfaces that i) directly interact with the environment through their embedded sensors, ii) make decisions and take actions, in a distributed way, in order to optimize the wave transformations that they apply to the radio waves, and iii) modify the radio waves based on the subsequent response from the environment. We refer to this process as environmental AI. \\

It is a known fact, however, that supervised machine learning methods require massive amounts of data that is difficult to gather in resource-constrained systems, or that is just not available in many application fields \cite{NEXUS}. Thinking of applying, on the other hand, reinforcement learning methods, it may take a very long time before the system converges to a stable and optimal operating point. In wireless networks, which are highly dynamic in nature, the system may not converge within the coherence time of the environment because of the well-known exploitation-exploration dilemma of reinforcement learning methods. There is, therefore, the compelling need of developing machine learning algorithms that can be optimized and designed by using small amount of data and that can optimally converge in a time much shorter than the coherence time of the wireless environment. \\

In \cite{152}, we have suggested and proved with some preliminary but promising results that transfer learning is a suitable approach in order to reduce the amount of data for system optimization. Transfer learning is a method that allows us to transfer the knowledge that is used in a given context to execute a given task, into a different but related context to execute another task \cite{149}. The approach that we have proposed in \cite{152} consists of combining together model-based and data-driven optimization methods. The idea is to exploit prior knowledge of the system based on mathematical models as the initialization point from which machine learning methods start interacting with the environment for system optimization. The rationale of the approach lies in the fact that the initial network status obtained from a model embeds many of the most important features of the actual system, and, therefore, it will take less time and data for machine learning methods to converge towards the optimal operating point. The results illustrated in \cite{152} in the context of optimizing the deployment of a cellular network are based on deep neural networks, and show promising results. However, making transfer learning work in wireless networks is not an easy task, since it is not guaranteed that the refinement from the initial state obtained from a model will lead to an optimized system that yields the same performance as a system that is optimized by using only a large amount of data. How to efficiently correct the mismatch between the model and the actual system with few empirical data, and to make the transfer of features positive, i.e., effective, is an open and challenging issue in transfer learning for wireless applications. \\

We think, in addition, that the future of wireless networks may be towards the realization of intelligent reconfigurable meta-surfaces with memory and computing power where machine learning is executed directly on the meta-surfaces by leveraging federated learning concepts \cite{152}, \cite{Federated}. The broad range of AI chipsets ranging from cloud AI to on-device AI, in fact, enable this opportunity \cite{Debbah}. \\ \\

\section{Concluding Discussion: Potential Impact}
The societal and economic impact of smart radio environments can be radical and profound \cite{21}. The vision of coating every environmental object with an intelligent reconfigurable meta-surface constitutes a transformative wireless future: Those objects that, ever since, have been perceived as an unintentional adversary to wireless communications are turned into programmable entities that help making communications and information processing more reliable and efficient. As an example: Smart radio environments can capitalize on the reflections of waves to make the received signal stronger, which has major benefits in virtual reality applications \cite{10}, and can be an enabler for reducing the transmit power in sensitive environments, e.g., hospitals, airplanes \cite{176}, where the cabin and the walls and ceilings can be coated with reconfigurable meta-surfaces. Also, we can think of smart cities, where ``smart'' encompasses the environment as well. \\

Smart radio environments largely expand the concept of network softwarization from the logical domain into the physical domain: The radio environment itself is viewed as a software entity, which can be remotely programmed, configured, and optimized. The concept of smart radio environments is not restricted to enhancing wireless communications, but is aimed at introducing a truly distributed intelligent wireless communications, sensing, and computing platform that interconnects the physical and digital worlds. Therefore, the expected impact of smart radio environments goes beyond wireless and embraces other fields of science, which include physics, computer science, and machine learning. \\

Smart radio environments constitute, in addition, an enabling sensing platform for interconnecting the physical and digital world. By recycling, e.g., the reflections of radio waves and embedding the data of sensors into them at a zero energy cost, the potential impact of smart radio environments is beyond communications. Imagine a smart radio space where the walls of rooms are coated with sensing meta-surfaces that monitor the health status of people. This will allow us to develop a truly pervasive and preventive e-health system. Imagine to coat with sensing meta-surfaces the bricks with which the kids play. This will allow us to discern how they put the bricks together, to infer their mental development, and how the human brain works \cite{PengyuZhang}. Imagine a smart home that learns our habits and configures the appliances around us as we desire, or that just monitors the network to understand its behavior and to prevent failures happening in the future \cite{12}. This will definitely improve our quality of life and will help us design better networks. Due to the large amount of energy that these three applications need, this vision is impossible with current technologies. It can be realized, on the other hand, by leveraging the concept of smart radio environments. \\ \\

In this paper, we have put forth a new communication-theoretic model for the analysis and optimization of smart radio environments, which explicitly accounts for the re-configurability of the radio environment via intelligent and reconfigurable meta-surfaces. Major research issues need, however, to be solved in order to make the vision of smart radio environments are reality:\\
\begin{itemize}
  \item How to integrate the reconfigurable meta-surfaces into wireless networks? \\
  \item What are the ultimate performance limits of wireless networks in the presence of reconfigurable meta-surfaces? \\
  \item How to attain such performance limits in practice? \\ \\
\end{itemize}

We hope that our newly introduced communication-theoretic model will motivate other researchers to develop the communication-theoretic and algorithmic foundation of smart radio environments empowered by intelligent and reconfigurable meta-surfaces. It is worth mentioning that, in fact, NTT DoCoMo and Metawave have recently run some experimental tests related to this technology, by using innovative 5G equipment provided by Ericsson and Intel \cite{DoCoMo-Metawave}. \\

As Marconi said many years ago, ``\textit{It is dangerous to put limits on wireless...}''.\\ \\

\begin{backmatter}

%%  After submission of the .TEX file,                     %%
%%  you will be prompted to submit your .BBL file.         %%

% if your bibliography is in bibtex format, use those commands:
\bibliographystyle{bmc-mathphys} % Style BST file (bmc-mathphys, vancouver, spbasic).
%\bibliography{bmc_article}      % Bibliography file (usually '*.bib' )

%%%%%%%%%%%%%%%%%%%%%%%%%%%%%%%%%%%
%%                               %%
%% Figures                       %%
%%                               %%
%% NB: this is for captions and  %%
%% Titles. All graphics must be  %%
%% submitted separately and NOT  %%
%% included in the Tex document  %%
%%                               %%
%%%%%%%%%%%%%%%%%%%%%%%%%%%%%%%%%%%

%%
%% Do not use \listoffigures as most will included as separate file

\end{backmatter}
\end{document}